\documentclass[11pt]{article}

\begin{document}

\date{}
\title{\textbf{Basic properties of Fedosov and Riemannian supermanifolds}}
\author{\textsc{M.~Asorey}${}^{a}$, 
\textsc{B.~Geyer}${}^{b}$, 
\textsc{P.M.~Lavrov}${}^{a,c}$ 
and
\textsc{O.V.~Radchenko}${}^{c}$ 
\\\\
${}^{a}$\textit{Departamento de F\'{\i}sica Te\'{o}rica, 
Facultad de Ciencias Universidad de Zaragoza,}\\
\textit{50009 Zaragoza, Spain}\\
${}^{b}$\textit{Institute of Theoretical Physics,
Leipzig University,}\\
\textit{D-04109 Leipzig, Germany}\\
${}^{c}$\textit{Department of Mathematical Analysis,
Tomsk State Pedagogical University,}\\
\textit{634041 Tomsk, Russia}}
\maketitle

\begin{quotation}
We discuss some differences in the properties of both even and odd
 Fedosov and Riemannian supermanifolds.
\end{quotation}

\vspace{.5cm}

A {\em Fedosov supermanifold} $(M,\omega,\Gamma)$ is defined as a symplectic
supermanifold $(M,\omega)$ equipped with a symmetric connection
$\Gamma$ (or covariant derivative $\nabla$) compatible with a given
symplectic structure $\omega$: $\omega\nabla=0$.\footnote{We use
conventions and definitions adopted in \cite{gl,lr,al}.} In local
coordinates $\{x^i\},\; \epsilon(x^i)=\epsilon_i$, on the
supermanifold $M$ the symplectic structure is
$\omega=\omega_{ij}dx^j\wedge dx^i,\quad
\omega_{ij}=-(-1)^{\epsilon_i\epsilon_j}\omega_{ji}$ (Grassmann
parity of the symplectic structure, $\epsilon(\omega),$ is equal to
0 for even structure and 1 for odd structure) and the compatibility
condition is $\omega_{ij}\nabla_k=\omega_{ij,k}-\Gamma_{ijk}+
\Gamma_{jik}(-1)^{\epsilon_i\epsilon_j}=0$ where
$\Gamma_{ijk}=\omega_{in}\Gamma^n_{\;jk},\quad
\epsilon(\Gamma_{ijk})=\epsilon(\omega)+
\epsilon_i+\epsilon_j+\epsilon_k$ and $\Gamma^i_{\;jk}$ are
components of the connection $\Gamma$. Notice that for a given
symplectic structure $\omega$ there exists a large family of
connections satisfying the compatibility condition.

The curvature tensor field $R^i_{\;mjk}$ is defined in terms  of the
commutator of covariant derivatives, $[\nabla_i,\nabla_j]=
\nabla_i\nabla_j-(-1)^{\epsilon_i\epsilon_j}\nabla_j\nabla_i$, whose
action on a vector field $T^i$ is $
T^i[\nabla_j,\nabla_k]=-(-1)^{\epsilon_m(\epsilon_i+1)}
T^mR^i_{\;mjk}$.

It is convenient to describe the basic properties of Fedosov
supermanifolds in terms of the symplectic curvature tensor  $
R_{ijkl}=\omega_{in}R^n_{\;\;jkl},\quad
\epsilon(R_{ijkl})=\epsilon(\omega)+\epsilon_i+
\epsilon_j+\epsilon_k+\epsilon_l, $.  This tensor obeys the symmetry
properties
\begin{eqnarray}
\label{Rans} R_{ijkl}=-(-1)^{\epsilon_k\epsilon_l}R_{ijlk}\,, \quad
R_{ijkl}=(-1)^{\epsilon_i\epsilon_j}R_{jikl}\,,
\end{eqnarray}
and satisfies the  Jacobi identity
\begin{eqnarray}
\label{Rjac0} (-1)^{\epsilon_j\epsilon_l}R_{ijkl}
+(-1)^{\epsilon_l\epsilon_k}R_{iljk}
+(-1)^{\epsilon_k\epsilon_j}R_{iklj}=0\,,
\end{eqnarray}
the Bianchi identity
\begin{eqnarray}
\label{Rjac1} (-1)^{\epsilon_k\epsilon_m}R_{ijkl;m}
+(-1)^{\epsilon_l\epsilon_m}R_{ijmk;l}
+(-1)^{\epsilon_k\epsilon_l}R_{ijlm;k}=0\,,
\end{eqnarray}
and the special symplectic identity
\begin{eqnarray}
\label{Rjac2} R_{ijkl}
+(-1)^{\epsilon_l(\epsilon_i+\epsilon_k+\epsilon_j)}R_{lijk}
+(-1)^{(\epsilon_k+\epsilon_l)(\epsilon_i+\epsilon_j)} R_{klij}+
(-1)^{\epsilon_i(\epsilon_j+\epsilon_l+\epsilon_k)}R_{jkli}=0.
\end{eqnarray}
We see that there are no formal differences in the properties of even
and odd Fedosov supermanifolds on the level of symplectic curvature
tensor. With  the curvature tensor, $R_{ijkl}$, and the inverse
tensor field $\omega^{ij}$
 ($\omega^{ij}= -(-1)^{\epsilon(\omega)+\epsilon_i\epsilon_j}\omega^{ji}$)
of the symplectic structure $\omega_{ij}$,
 one can construct the only  tensor field of type $(0,2)$,
\begin{eqnarray}
K_{ij}= \omega^{kn}R_{nikj}
 (-1)^{\epsilon_i\epsilon_k+(\epsilon(\omega)+1)(\epsilon_k+\epsilon_n)}
 \;=\;R^k_{\;\;ikj}\;(-1)^{\epsilon_k(\epsilon_i+1)},\quad
 \epsilon(K_{ij})=\epsilon_i+\epsilon_j.
 \end{eqnarray}
This tensor satisfies  the relations \cite{gl}
\begin{eqnarray}
\label{Rl3}
[1+(-1)^{\epsilon(\omega)}](K_{ij}-(-1)^{\epsilon_i\epsilon_j}K_{ji})=0\,,
\end{eqnarray}
and is called the Ricci tensor. In the even case this tensor is
symmetric whereas  in the odd case there are not restrictions on its
(generalized) symmetry properties. The scalar curvature tensor $K$
is defined by the formula $
K=\omega^{ji}K_{ij}(-1)^{\epsilon_i+\epsilon_j}$. From the symmetry
properties of $R_{ijkl}$, it follows that
\begin{eqnarray}
\label{Rsc1} [1+(-1)^{\epsilon(\omega)}]K=0.
\end{eqnarray}
Therefore  as in the case of Fedosov manifolds \cite{fm}, even
Fedosov  supermanifolds have vanishing scalar curvature $K$.
However, for odd Fedosov supermanifolds this curvature  is, in
general, not vanishing. This fact was quite recently used in Ref.
\cite{BB} to  generalize the BV formalism \cite{bv}.
\medskip

A {\em Riemannian supermanifold} $(M,g,\Gamma)$ is defined as a metric
supermanifold $(M,g)$ equipped with a symmetric connection $\Gamma$
(or covariant derivative $\nabla$) compatible with a given metric
structure $g$: $g\nabla=0$. In local coordinates on the
supermanifold $M$ the metric structure is $g=g_{ij}dx^j dx^i,\quad
g_{ij}=(-1)^{\epsilon_i\epsilon_j}g_{ji}$ (Grassmann parity of the
metric structure, $\epsilon(g),$ is equal to 0 for even structure
and 1 for odd structure) and the compatibility condition is
$g_{ij}\nabla_k=g_{ij,k}-\Gamma_{ijk}-
\Gamma_{jik}(-1)^{\epsilon_i\epsilon_j}=0$ where
$\Gamma_{ijk}=g_{in}\Gamma^n_{\;jk},\quad
\epsilon(\Gamma_{ijk})=\epsilon(g)+
\epsilon_i+\epsilon_j+\epsilon_k$ and $\Gamma^i_{\;jk}$ are
components of the connection $\Gamma$. Notice that for a given
metric structure $g$ there exists the unique symmetric connection
$\Gamma^i_{\;jk}$ which is compatible with a given metric structure,
\begin{eqnarray}\label{gDelta}
\Gamma^l_{\;ki}=\frac{1}{2}g^{lj}\Big(g_{ij,k}
(-1)^{\epsilon_k\epsilon_i} +g_{jk,i}(-1)^{\epsilon_i\epsilon_j}-
g_{ki,j}(-1)^{\epsilon_k\epsilon_j}\Big)(-1)^
{\epsilon_j\epsilon_i+\epsilon_j+\epsilon(g)(\epsilon_j+\epsilon_l)},
\end{eqnarray}
where  $g^{ij}$ is the inverse tensor field of the metric $g_{ij}$
($ g^{ij}=(-1)^{\epsilon(g)+\epsilon_i\epsilon_j}g^{ji},\quad
\epsilon(g^{ij})=\epsilon(g)+\epsilon_i+\epsilon_j).$

 The curvature tensor ${\cal R}_{ijkl}=g_{in}{\cal
R}^n_{\;jkl}$ $(\epsilon({\cal R}_{ijkl})=\epsilon(g)+\epsilon_i+
\epsilon_j+\epsilon_k+\epsilon_l)$, obeys the symmetry properties
\begin{eqnarray}
\label{Rans1} {\cal R}_{ijkl}=-(-1)^{\epsilon_k\epsilon_l}{\cal
R}_{ijlk}, \quad {\cal R}_{ijkl}=-(-1)^{\epsilon_i\epsilon_j}{\cal
R}_{jikl},\quad {\cal R}_{ijkl}={\cal
R}_{klij}(-1)^{(\epsilon_i+\epsilon_j) (\epsilon_k+\epsilon_l)}
\end{eqnarray}
and satisfies the Jacobi identity (\ref{Rjac0}) and the Bianchi
identity (\ref{Rjac1}). Again we find that on the level of the curvature
tensor there are no differences in the basic properties of even and odd
Riemannian supermanifolds.

From the  curvature tensor ${\cal R}_{ijkl}$ and the inverse tensor
field $g^{ij}$ of the metric $g_{ij}$ one can define the only tensor
field of type $(0,2)$:
\begin{eqnarray}
 \label{3tf2} {\cal R}_{ij}={\cal
R}^k_{\;\;ikj}(-1)^{\epsilon_k(\epsilon_i+1)}= g^{kn}{\cal R}_{nikj}
(-1)^{(\epsilon_k+\epsilon_n)(\epsilon(g)+1)+\epsilon_i\epsilon_k},\quad
\epsilon({\cal R}_{ij})=\epsilon_i+ \epsilon_j.
\end{eqnarray}
It is the generalized   Ricci tensor with the following symmetry
properties
\begin{eqnarray}
R_{ij}=(-1)^{\epsilon(g)+\epsilon_i\epsilon_j}R_{ji}
\end{eqnarray}
depending on Riemannian supermanifolds to be even or odd. A further
contraction defines  the scalar curvature
\begin{eqnarray}
\label{Scalcur} {\cal R} = g^{ji}{\cal
R}_{ij}\;(-1)^{\epsilon_i+\epsilon_j},\quad \epsilon({\cal
R})=\epsilon(g)
\end{eqnarray}
which, in general, is not equal to zero. Notice that for an odd
metric structure the scalar curvature tensor squared is identically
equal to zero, ${\cal R}^2=0$.

From the Bianchi identity one can deduce  the following relation
between the  scalar curvature  and the Ricci tensor
\begin{eqnarray}
\label{RicciscR1} {\cal R}_{,i}=[1+(-1)^{\epsilon(g)}]{\cal
R}^{j}_{\;\;i;j} (-1)^{\epsilon_j(\epsilon_i+1)},
\end{eqnarray}
which in the even case is nothing but ${\cal R}_{,i}=2{\cal
R}^{j}_{\;\;i;j} (-1)^{\epsilon_j(\epsilon_i+1)}$ , i.e. the
supersymmetric generalization of known relation in Riemannian
geometry \cite{Eisenhart}. In the odd case ${\cal R}_{,i}=0$.
Therefore  odd Riemann supermanifolds have constant scalar
curvature, ${\cal R}= const$.
\\


\end{document}